# Acoustic radiation force and torque on a viscous fluid cylindrical particle nearby a planar rigid wall

F.G. Mitri[a]

*Chevron, Area 52 Technology – ETC, Santa Fe, New Mexico 87508, USA*

**Abstract**

This work presents a comprehensive analytical formalism for the modal series expansions of the acoustic radiation force and radiation torque experienced by a fluid viscous cylindrical object of arbitrary geometrical cross-section placed near a planar rigid wall. A plane progressive wave field with an arbitrary angle of incidence propagating in an inviscid fluid is assumed. The formalism developed here utilizes the features of the modal expansion method with boundary matching in cylindrical coordinates, the method of images and the translational addition theorem. Initially, an *effective* field incident on the particle (including the principal incident field, the reflected wave-field from the boundary and the scattered field from the image object) is defined. Subsequently, the incident effective field is utilized in conjunction with the scattered one from the object, to obtain closed-form expansions for the longitudinal and transversal force components, in addition to the axial torque component, based on the scattering in the far-field. The obtained solutions involve the angle of incidence, the modal coefficients of the scatterer and its image, and the particle-wall distance. Computations for the non-dimensional force and torque functions for a viscous fluid circular cylindrical cross-section are considered, and calculations exemplify the analysis with emphasis on changing the particle size, the particle-wall distance and the angle of incidence. It is found that the object experiences either an attractive or repulsive force directed toward or away from the boundary. Moreover, it reverses its rotation around its center of mass depending on its size, particle-wall distance and angle of incidence. Furthermore, radiation force and torque singularities occur (i.e., zero force and torque) such that the cylinder becomes irresponsive to the linear and angular momenta transfer yielded by the effective field. The results lead to an improved understanding of the force and radiation torque behaviors for a viscous particle in the field of plane progressive waves nearby a boundary, which is a commonly encountered situation in acoustofluidics, contrast agents in biomedicine, and fluid dynamics to name a few examples. The exact formalism presented here using the multipole expansion method, which is valid for any frequency range, could be used to validate other approaches using purely numerical methods, or frequency-limiting approximate models such as the Rayleigh and the Kirchhoff regimes.

## 1. Introduction

When evaluating the radiation force and radiation torque on a particle located nearby a boundary [1] or a chamber wall [2], it is crucial to take into account the multiple interactions, i.e. scatterings [3] and reflections. Therefore, a need to predict numerically the acoustic forces [1,2,4] and torque accurately near a chamber walls arises. This effort has been originally presented in [1] for an elastic spherical contrast agent shell nearby a porous flat boundary.

Several investigations were previously developed for a cylindrical particle in an unbounded fluid [5-14], where some recent analyses (not limited to a particular range of frequencies) considered the elliptical geometry [15-18]. Nonetheless, those formalisms cannot be applied to a cylindrical particle nearby a boundary, and it is important to develop an improved methodology taking the multiple reflections/scattering effects between the wall and the particle into account.

The aim of this investigation is therefore directed toward the development of a novel analytical formalism for the acoustic radiation force and torque on a particle of arbitrary geometrical cross-section (in 2D) valid for any scattering regime (i.e., Rayleigh, Mie or geometrical/ray acoustics), which is yet to be developed in the scientific literature. Particularly, the rigorous formalism accounts for the multiple reflections/scattering effects occurring between the particle and the rigid boundary by means of the modal expansion method in cylindrical coordinates [19], the method of images [20,21] and the translational addition theorem [22].

The procedure to obtain the acoustic radiation force and radiation torque consists on the integration of the time-averaged Brillouin's stress tensor [23,24] and its moment [25,26], respectively. The procedure uses an analysis of the scattering in the far-field, without any approximation in the evaluation of the physical observables. An effective acoustic velocity potential field incident on the particle is defined, which includes the primary incident field, the reflected waves from the boundary as well as the scattered field from the image object, is utilized with the scattered field from the object to obtain the radiation force and radiation torque functions. The present investigation is substantiated by numerical computations for a lossy fluid cylindrical particle of circular cross-section illuminated by plane travelling waves with arbitrary incidence. The numerical examples are chosen to illustrate possible practical scenarios with emphases on the distance of the particle from the boundary, the dimensionless size parameter, and the incidence angle of the primary waves.

---

[a] Electronic mail: F.G.Mitri@ieee.org

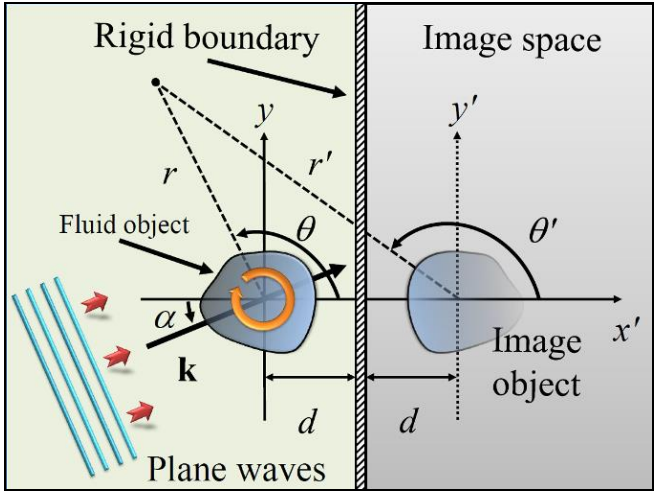

FIG. 1. Sketch representing a cylindrical fluid object of arbitrary cross-section (in 2D) submerged in an unbounded non-viscous fluid. Its center of mass is located at a distance $d$ from a rigid planar wall. Acoustic plane progressive wave illumination with an arbitrary incidence angle $\alpha$ is assumed. The rotating arrow on the fluid object surface represents the induced spinning (counter-clockwise or clockwise) caused by the axial $z$-component of acoustic radiation torque vector.

The mathematical model and its related physical results may be important for applications involving the numerical predictions of the acoustic radiation force and radiation torque induced by plane waves of arbitrary incidence on a particle located nearby a boundary, using *exact* partial-wave series expansions (PWSEs). The results can assist in validating numerical tools using the finite-element or boundary-element methods (FEM/BEM) [27,28], and other techniques. Moreover, the 2D cylindrical particle geometry serves as a benchmark solution where the anticipated physical effects occurring in 2D are also expected to exist for a 3D particle close to a boundary, and the present model should assist to further develop an improved formalism for the radiation force and radiation torque on a spherical particle located nearby a wall [1] with arbitrary incidence.

## 2. Method

### 2.1. Plane waves scattering by a fluid cylindrical particle of arbitrary geometry in 2D near a rigid flat boundary

Consider a particle with arbitrary geometrical cross-section, subjected to a continuous-wave field of harmonic plane travelling waves propagating in an inviscid fluid. The incident wave-field has an angle of incidence $\alpha$ with respect to the horizontal plane as shown in Fig. 1. The expression for the incident velocity potential is given as

$$\Phi_{\text{inc}} = \phi_0 e^{i(\mathbf{k}\cdot\mathbf{r}-\omega t)}, \tag{1}$$

where $\phi_0$ is the amplitude, $\omega$ is the angular frequency, $\mathbf{k}$ is the wave-vector and $\mathbf{r}$ is the vector position.

In the cylindrical system of coordinates $(r,\theta)$, Eq.(1) is expressed as

$$\begin{aligned}\Phi_{\text{inc}}(r,\theta,t) &= \phi_0 e^{-i\omega t} e^{ikr\cos(\theta-\alpha)} \\ &= \phi_0 e^{-i\omega t}\sum_{n=-\infty}^{+\infty} i^n e^{-in\alpha} J_n(kr) e^{in\theta},\end{aligned} \tag{2}$$

where $k = \omega/c$ is the wavenumber, $c$ is the speed of sound in the medium of wave propagation, and $J_n(\cdot)$ is the cylindrical Bessel function of the first kind.

Due to the existence of the boundary, reflection of the incident (primary) field occurs, which is in turn incident on the fluid object. The velocity potential field for the reflected waves, incident upon the fluid object, is expressed as

$$\begin{aligned}\Phi_R(r,\theta,t) &= \phi_0 e^{-i\omega t} e^{ikr\cos(\theta-\pi+\alpha)} e^{2ikd\cos\alpha} \\ &= \phi_0 e^{-i\omega t} e^{2ikd\cos\alpha}\sum_{n=-\infty}^{+\infty} i^n e^{-in(\pi-\alpha)} J_n(kr) e^{in\theta},\end{aligned} \tag{3}$$

where $d$ is the distance from the center of mass of the particle to the boundary along the direction $\theta = 0$ (Fig. 1).

The scattered velocity potential field resulting from the interaction of the primary incident plane waves with the object is expressed as,

$$\Phi_{\text{sca}}^{\text{object}}(r,\theta,t) = \phi_0 e^{-i\omega t}\sum_{n=-\infty}^{+\infty} C_n H_n^{(1)}(kr) e^{in\theta}, \tag{4}$$

where $H_n^{(1)}(\cdot)$ is the cylindrical Hankel function of the first kind of order $n$, and $C_n$ is the modal coefficient to be determined subsequently.

Since the object is sound penetrable, the internal velocity potential field for the compressional waves propagating in its core is written as

$$\Phi_{\text{int}}^{\text{object}}(r,\theta,t) = \phi_0 e^{-i\omega t}\sum_{n=-\infty}^{+\infty} C_n^{\text{int}} J_n(\kappa_L r) e^{in\theta}, \tag{5}$$

where $\kappa_L$ is the wavenumber corresponding to the longitudinal/compressional waves propagating inside the core fluid material. For a viscous fluid medium, $\kappa_L$ is a complex number, accounting for sound absorption. $C_n^{\text{int}}$ is the expansion coefficient for the internal waves.

The method of images is now applied so that the scattering from the flat rigid boundary can be replaced by the scattering from the image object (mirrored by the wall). Therefore, the scattered velocity potential contributed by the image object is

given as

$$\Phi_{sca}^{image}(r',\theta',t) = \phi_0 e^{-i\omega t} \sum_{n=-\infty}^{+\infty} D_n H_n^{(1)}(kr') e^{in\theta'}, \quad (6)$$

where $D_n$ is the modal coefficient of the image object.

In addition, the internal velocity potential field internal to the image object is expressed as

$$\Phi_{int}^{image}(r',\theta',t) = \phi_0 e^{-i\omega t} \sum_{n=-\infty}^{+\infty} D_n^{int} J_n(\kappa_L r') e^{in\theta'}, \quad (7)$$

where $D_n^{int}$ is the expansion coefficient for the internal waves of the image object.

Eqs.(4)-(7) are valid for any particle of arbitrary geometry (in 2D) located nearby a flat rigid boundary and its image. The expansion series (written in terms of cylindrical wave functions) were utilized earlier in investigations dealing with elastic wave [29-31] and acoustic scattering [15,17,32,33] using the T-matrix [34]. (Other discussions for non-circular cylinders can be found in [15-18,35,36]).

Before applying the appropriate boundary conditions to obtain the modal coefficients, the expressions for the velocity potential fields must be modified such that they are all referred to the same coordinates system. The adequate boundary conditions of continuity must be satisfied, either at the planar rigid boundary surface (which were used previously in the context of the multiple acoustic scattering [37,38] from a rigid boundary [39,40]), or at the surface of the particle and its corresponding image in both systems of coordinates [41]. Both methods are commensurate with the same result. Using the latter approach, the translational addition theorem is used to express the corresponding PWSE in a particular system as a series in the other system.

Near the object and its image, the following expressions hold [42],

$$H_n^{(1)}(kr') e^{in\theta'} = \sum_{m=-\infty}^{+\infty} J_m(kr) H_{m-n}^{(1)}(2kd) e^{im\theta}, \quad r<2d, \quad (8)$$

whereas,

$$H_n^{(1)}(kr) e^{in\theta} = \sum_{m=-\infty}^{+\infty} J_m(kr') H_{n-m}^{(1)}(2kd) e^{im\theta'}, \quad r'<2d. \quad (9)$$

Subsequently, in the fluid host medium surrounding the object, the total velocity potential is written as,

$$\Phi_{tot}(r,\theta,t)\big|_{r<2d}$$
$$= \Phi_{inc}(r,\theta,t) + \Phi_R(r,\theta,t) \quad (10)$$
$$+ \Phi_{sca}^{object}(r,\theta,t) + \Phi_{sca}^{image}(r,\theta,t)\big|_{r<2d},$$

where,

$$\Phi_{sca}^{image}(r,\theta,t)\big|_{r<2d}$$
$$= \phi_0 e^{-i\omega t} \sum_{n=-\infty}^{+\infty} \left( \sum_{m=-\infty}^{+\infty} D_m H_{n-m}^{(1)}(2kd) \right) J_n(kr) e^{in\theta}. \quad (11)$$

Following the same procedure, the expansion series for the total velocity potential field in the primed system of coordinates is obtained as,

$$\Phi_{tot}(r',\theta',t)\big|_{r'<2d}$$
$$= \Phi_{inc}(r',\theta',t) + \Phi_R(r',\theta',t) \quad (12)$$
$$+ \Phi_{sca}^{object}(r',\theta',t)\big|_{r'<2d} + \Phi_{sca}^{image}(r',\theta',t),$$

where,

$$\Phi_{inc}(r',\theta',t) = \phi_0 e^{-i\omega t} \sum_{n=-\infty}^{+\infty} i^n e^{-in(\pi-\alpha)} J_n(kr') e^{in\theta'}, \quad (13)$$

$$\Phi_R(r',\theta',t) = \phi_0 e^{-i\omega t} e^{2ikd\cos\alpha} \sum_{n=-\infty}^{+\infty} i^n e^{-in\alpha} J_n(kr') e^{in\theta'}, \quad (14)$$

and,

$$\Phi_{sca}^{object}(r',\theta',t)\big|_{r'<2d} = \phi_0 e^{-i\omega t} \sum_{n=-\infty}^{+\infty} \left( \sum_{m=-\infty}^{+\infty} C_m H_{m-n}^{(1)}(2kd) \right) J_n(kr') e^{in\theta'}. \quad (15)$$

Since the particle is a sound penetrable fluid, the following conditions of continuity should be imposed in both systems of coordinates (i.e. object + image), such that,

○   $\Phi_{tot}(r,\theta,t)\big|_{r=A_\theta} = \Phi_{int}^{object}(r,\theta,t)\big|_{r=A_\theta},$   (16)

○   $\left(\dfrac{1}{\rho}\right)\nabla\Phi_{tot}(r,\theta,t)\cdot\mathbf{n}\big|_{r=A_\theta} = \left(\dfrac{1}{\rho_f}\right)\nabla\Phi_{int}^{object}(r,\theta,t)\cdot\mathbf{n}\big|_{r=A_\theta},$   (17)

○   $\Phi_{tot}(r',\theta',t)\big|_{r'=A_\theta} = \Phi_{int}^{image}(r',\theta',t)\big|_{r'=A_\theta},$   (18)

○   $\left(\dfrac{1}{\rho}\right)\nabla\Phi_{tot}(r',\theta',t)\cdot\mathbf{n}\big|_{r'=A_\theta} = \left(\dfrac{1}{\rho_f}\right)\nabla\Phi_{int}^{image}(r',\theta',t)\cdot\mathbf{n}\big|_{r'=A_\theta},$   (19)

where $\rho$ and $\rho_f$ are the mass densities of the host fluid medium and fluid particle, respectively. The parameters $\mathbf{n}$ and $A_\theta$ are the normal vector and the surface shape function of the particle of arbitrary geometry [15-18], respectively.

## 2.2. Radiation force and radiation torque expressions

The series expansions for the longitudinal and transversal vector force components are attained after defining an incident effective velocity potential field, which results from combining the principal incident field with the reflected waves from the flat wall as well as the scattered waves from the image object [see Eq.(20)]. This improved definition allows a simplification of the problem in hand, such that the determination of the components is equivalent to the sole object, illuminated by an effective field, which accounts for the multiple scatterings, re-scattered waves and particle-wall interactions.

Recognizing that the (cylindrically-outgoing) scattered waves cannot be concurrently incident upon the object surface, the incident effective velocity potential field in the system of coordinates $(r,\theta)$ is determined from Eq.(10) such that,

$$\Phi_{inc}^{eff}(r,\theta,t) = \Phi_{tot}(r,\theta,t)\Big|_{r<2d} - \Phi_{sca}^{object}(r,\theta,t)$$
$$= \Phi_{inc}(r,\theta,t) + \Phi_R(r,\theta,t) + \Phi_{sca}^{image}(r,\theta,t)$$
$$= \phi_0 e^{-i\omega t}\left[\sum_{n=-\infty}^{+\infty} i^n \left(e^{-in\alpha} + e^{-in(\pi-\alpha)+2ikd\cos\alpha}\right) J_n(kr) e^{in\theta}\right.$$
$$\left. + \sum_{n=-\infty}^{+\infty}\left(\sum_{m=-\infty}^{+\infty} D_m H_{n-m}^{(1)}(2kd)\right) J_n(kr) e^{in\theta}\right].$$
(20)

After determination of the effective velocity potential field, the formulation is extended to derive the analytical components (i.e., longitudinal and transversal) of the radiation force based on the far-field scattering approach [23,43-45]. The mathematical functions are expressed, respectively, by the following approximations; $J_n(kr) \underset{kr\to\infty}{\approx} \sqrt{\frac{2}{\pi kr}} \cos\left(kr - \frac{n\pi}{2} - \frac{\pi}{4}\right)$ and $H_n^{(1)}(kr) \underset{kr\to\infty}{\approx} \sqrt{\frac{2}{\pi kr}} e^{i\left(kr - \frac{n\pi}{2} - \frac{\pi}{4}\right)}$, and the radiation (particle-wall) force series expansions are obtained based on integrating Brillouin's stress over a surface with a large radius [23,43,46] surrounding the object. An alternative method consists of integrating Brillouin's stress tensor over the surface of the probed particle using the approach of the near-field scattering. Notice, however, that in an inviscid host fluid, these two approaches are equivalent [45,47]. Nonetheless, the benefit of using the far-field scattering lies in achieving fewer algebraic manipulations.

The expression for the force vector used previously [45] can be applied such that,

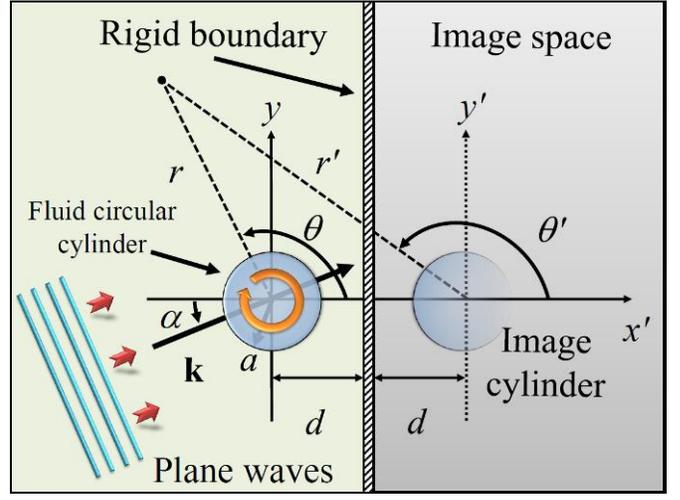

FIG. 2. The same as in Fig. 1, but the cylinder has a circular cross-section of radius $a$.

$$\langle\mathbf{F}\rangle \underset{kr\to\infty}{=} \frac{1}{2}\rho k^2 \int_0^{2\pi} \Re\{\Phi_{is}\} d\mathbf{S},$$
(21)

where $\Re\{\cdot\}$ denotes the real part of a complex number, $\Phi_{is} \underset{kr\to\infty}{=} \Phi_{sca}^{object*}\left[(i/k)\partial_r\Phi_{inc}^{eff} - \Phi_{inc}^{eff} - \Phi_{sca}^{object}\right]$, and $d\mathbf{S} = dS\,\mathbf{e}_r$, where $dS = r\,d\theta$ is differential element of a surface $S$ of unit-length enclosing the particle. The normal unit vector (pointing outwardly to the surface) is $\mathbf{e}_r = \cos\theta\,\mathbf{e}_x + \sin\theta\,\mathbf{e}_y$, where $\mathbf{e}_x$ and $\mathbf{e}_y$ are the unit vectors, the symbol $\langle\cdot\rangle$ denotes time-averaging, and the superscript * denotes the conjugate of a complex number.

The force vector consists of a longitudinal (along the $x$-direction) and a transversal (along the $y$-direction) component. Those are defined as

$$\begin{Bmatrix}F_x\\F_y\end{Bmatrix} = \langle\mathbf{F}\rangle \cdot \begin{Bmatrix}\mathbf{e}_x\\\mathbf{e}_y\end{Bmatrix}$$
$$= \begin{Bmatrix}Y_x\\Y_y\end{Bmatrix} S_c E_0,$$
(22)

where $S_c$ is the cross-sectional surface of the particle, $E_0 = \frac{1}{2}\rho k^2 |\phi_0|^2$ is a representative energy density parameter, and $Y_x$ and $Y_y$ are the non-dimensional longitudinal and transversal radiation force functions, respectively.

Using the angular integrals,

$$\int_0^{2\pi} e^{i(n'-n)\theta} \begin{Bmatrix} \cos\theta \\ \sin\theta \end{Bmatrix} d\theta = \pi \begin{Bmatrix} (\delta_{n,n+1} + \delta_{n,n-1}) \\ i(\delta_{n,n+1} - \delta_{n,n-1}) \end{Bmatrix}, \quad (23)$$

where $\delta_{ij}$ is the Kronecker delta function, Eq.(21) gives the mathematical series for the longitudinal and transversal force components, respectively, as

$$Y_x = \frac{2}{kS_c}\Im\left\{\sum_{n=-\infty}^{+\infty}\left(i^n\left[e^{-in\alpha} + e^{-in(\pi-\alpha)+2ikd\cos\alpha}\right] + C_n + \sum_{m=-\infty}^{+\infty} D_m H_{n-m}(2kd)\right)(C_{n+1}^* - C_{n-1}^*)\right\}, \quad (24)$$

$$Y_y = -\frac{2}{kS_c}\Re\left\{\sum_{n=-\infty}^{+\infty}\left(i^n\left[e^{-in\alpha} + e^{-in(\pi-\alpha)+2ikd\cos\alpha}\right] + C_n + \sum_{m=-\infty}^{+\infty} D_m H_{n-m}(2kd)\right)(C_{n+1}^* + C_{n-1}^*)\right\}, \quad (25)$$

where $\Im\{\cdot\}$ designates the imaginary part of a complex number.

The closed-form partial-wave series expansion for the axial component (along the $z$-direction) of the radiation torque vector is also obtained as follows.

Using the general expression for the radiation torque vector [25] (which was also used in the context of a cylindrically-focused quasi-Gaussian beam incident on a viscoelastic cylinder [14], plane waves on an elliptical cylinder [18] and a pair of fluid cylinders [35]) given by,

$$\langle \mathbf{N} \rangle_{kr\to\infty} = -\rho \iint_S \langle \mathbf{v} \otimes (\mathbf{r} \times \mathbf{v}) \rangle \cdot d\mathbf{S}, \quad (26)$$

where the symbol $\otimes$ denotes a tensor product, and $\mathbf{v} = \nabla\left(\Phi_{\text{inc}}^{\text{eff}} + \Phi_{\text{sca}}^{\text{object}}\right)$, the sole non-vanishing axial radiation torque component [in the $z$-direction normal to the polar plane $(r,\theta)$] can be rewritten as,

$$N_z^{\text{rad}} = \langle \mathbf{N} \rangle \cdot \mathbf{e}_z,$$
$$= \frac{\rho}{2}\Im\left\{\iint_S \left(\partial_r \Phi_{\text{inc}}^{\text{eff}*} + \partial_r \Phi_{\text{sca}}^{\text{object}*}\right)\hat{L}_z\left(\Phi_{\text{inc}}^{\text{eff}} + \Phi_{\text{sca}}^{\text{object}}\right) dS\right\}, \quad (27)$$

where $\mathbf{e}_z$ is the unit vector along the $z$-axis, and $\hat{L}_z$ is the axial component of the angular momentum operator in polar coordinates given by,

$$\hat{L}_z = -i\frac{\partial}{\partial\theta}. \quad (28)$$

After some calculation using the property of the following angular integral,

$$\int_0^{2\pi} e^{i(n'-n)\theta} d\theta = 2\pi\delta_{n,n'}, \quad (29)$$

The axial radiation torque expression is obtained as,

$$N_z^{\text{rad}} = \tau_z V E_0, \quad (30)$$

where $V$ is the particle volume, and $\tau_z$ is the non-dimensional axial torque function, given by

$$\tau_z = -\frac{4}{k^2 V}\Re\left\{\sum_{n=-\infty}^{+\infty} n\, C_n^*\begin{pmatrix} i^n\left[e^{-in\alpha} + e^{-in(\pi-\alpha)+2ikd\cos\alpha}\right] + C_n \\ + \sum_{m=-\infty}^{+\infty} D_m H_{n-m}(2kd) \end{pmatrix}\right\}. \quad (31)$$

Eqs.(24),(25) and (31) are generalized series expansions, suitable for a 2D object of arbitrary geometry located nearby a flat boundary, and illuminated by a plane travelling wave-field with an arbitrary angle of incidence.

### 2.3. Circular fluid cylinder example

The example of a liquid circular cylinder of radius $a$, as shown in Fig. 2, is now considered. For the circular geometry, the boundary conditions given by Eqs. (16) and (17) at $r = a$ lead to,

$$i^{\ell}\left(e^{-i\ell\alpha} + e^{2ikd\cos\alpha}e^{-i\ell(\pi-\alpha)}\right)J_{\ell}(ka) + C_{\ell}H_{\ell}^{(1)}(ka)$$
$$+ J_{\ell}(ka)\sum_{n=-\infty}^{\infty}D_{n}H_{\ell-n}^{(1)}(2kd) = C_{\ell}^{\text{int}}J_{\ell}(\kappa_{L}a), \quad (32)$$

$$i^{\ell}\left(e^{-i\ell\alpha} + e^{2ikd\cos\alpha}e^{-i\ell(\pi-\alpha)}\right)J_{\ell}'(ka) + C_{\ell}H_{\ell}^{(1)'}(ka)$$
$$+ J_{\ell}'(ka)\sum_{n=-\infty}^{\infty}D_{n}H_{\ell-n}^{(1)}(2kd) = \left(\tfrac{\rho c}{\rho_{f}c_{f}}\right)C_{\ell}^{\text{int}}J_{\ell}'(\kappa_{L}a), \quad (33)$$

where the primes in Eq.(33) [and Eq.(35), below] denote the derivatives of the wave functions, and $c_f$ is the speed of sound of the compressional waves in the fluid cylinder.

Correspondingly, the application of the boundary conditions at the surface of the image cylinder leads to,

$$i^{\ell}\left(e^{2ikd\cos\alpha}e^{-i\ell\alpha} + e^{-i\ell(\pi-\alpha)}\right)J_{\ell}(ka) + D_{\ell}H_{\ell}^{(1)}(ka)$$
$$+ J_{\ell}(ka)\sum_{n=-\infty}^{\infty}C_{n}H_{n-\ell}^{(1)}(2kd) = D_{\ell}^{\text{int}}J_{\ell}(\kappa_{L}a), \quad (34)$$

$$i^{\ell}\left(e^{2ikd\cos\alpha}e^{-i\ell\alpha} + e^{-i\ell(\pi-\alpha)}\right)J_{\ell}'(ka) + D_{\ell}H_{\ell}^{(1)'}(ka)$$
$$+ J_{\ell}'(ka)\sum_{n=-\infty}^{\infty}C_{n}H_{n-\ell}^{(1)}(2kd) = \left(\tfrac{\rho c}{\rho_{f}c_{f}}\right)D_{\ell}^{\text{int}}J_{\ell}'(\kappa_{L}a). \quad (35)$$

The coupled system of Eqs.(32)-(35) can be calculated numerically by matrix inversion, so that the modal coefficients $C_n$ and $D_n$ can be determined. Then, the force and torque vector components can be computed after adequate truncation of the series, with a suitable limit that ensures convergence. The truncation limit in the series has been set to $N_{max}$ = [(max($ka$, $kd$)] + 35, which leads to adequate convergence and accuracy of the results with negligible truncation error in the order of ~$10^{-6}$.

### 3. Numerical examples and discussions

Numerical examples are considered for which a MATLAB program is developed to compute the expansion coefficients for the fluid viscous circular cylinder located nearby a rigid flat boundary, and subsequently compute the longitudinal and transversal force and the axial torque components (per-length). For the circular cylinder case, $S_c = 2a$ and $V = \pi a^2$ for a *unit-length* cylinder.

The parameters chosen for the viscous fluid circular cylinder mimic those of a red blood fluid material, pertinent to some bio-acoustofluidics applications. The mass density of the fluid circular cylinder particles is $\rho_f$ = 1099 kg/m$^3$, and the speed of sound for the compressional waves in its core is $c_f$ = 1631 m/s. The host liquid is water ($\rho$ = 1000 kg/m$^3$ and speed of sound $c$ = 1500 m/s). Inside the inner core cylinder material, the (complex) wavenumber for the compressional waves is expressed as $\kappa_L = \omega/c_f(1+i\gamma)$, where $\gamma (= 10^{-3})$ is a dimensionless absorption coefficient [48].

Computations for the longitudinal and transversal force components, in addition to the axial torque component are performed where the angle of incidence of the plane progressive wave field and the dimensionless distance are varied, respectively, in the chosen ranges $-90° \leq \alpha \leq 90°$ and $ka < kd \leq 15$, for fixed values of the size parameter $ka$ of the circular cylinder. First, the example of a Rayleigh circular viscous fluid cylinder near a boundary having $ka$ = 0.1 is considered. Panel (a) of Fig. 3 displays the plot for the longitudinal radiation force function, which exhibits periodic oscillations versus $kd$ as $\alpha$ varies in the range $-90° \leq \alpha \leq 90°$. These oscillations result from multiple reflections occurring between the circular cylinder and rigid wall, where they are maximal at $\alpha = 0$. As $|\alpha|$ increases, the multiple reflections become less pronounced due to geometrical consideration as the waves reflected from the boundary become less scattered by the particle. The longitudinal force components varies between positive and negative values depending on the dimensionless distance $kd$ and $\alpha$, suggesting a pushing or pulling effects towards the acoustic source. At the crossing point between positive and negative values, a longitudinal radiation force function singularity occurs where $Y_x = 0$. This effect occurs manifold as $kd$ and $\alpha$ vary, suggesting that the fluid viscous Rayleigh cylinder becomes unresponsive to the longitudinal force, and is only subjected to the transversal force component. Moreover, symmetry with respect to the axis $\alpha = 0$ is noted such that $Y_x(-\alpha) = Y_x(\alpha)$. Panel (b) shows the plot for the transversal radiation force function, which displays asymmetric values such that $Y_y(-\alpha) = -Y_y(\alpha)$ as $kd$ increases with respect to the axis $\alpha = 0$, where $Y_y$ vanishes therein due to symmetry considerations. In contrast to what has been observed for the longitudinal radiation force component in panel (a), as $|\alpha|$ increases, the effect of the multiple reflections on $Y_y$ becomes more pronounced as its amplitude displays larger oscillations as $|\alpha|$ approaches 90°. Panel (c) shows the plot for the axial radiation torque function, which also displays asymmetry with respect to the axis $\alpha = 0$, where it vanishes therein due to symmetry. Notice also that $\tau_z(-\alpha) = -\tau_z(\alpha)$. In panel (c), it is shown that $\tau_z$ varies between positive and negative values, suggesting that the viscous fluid circular cylinder is subjected to a radiation torque causing its rotation counter-clockwise or clockwise. Moreover, at the crossing point between positive and negative values of $\tau_z$, an axial radiation torque function singularity occurs where $\tau_z = 0$ such that there is no spinning around the center of mass of the cylinder; this suggests that the viscous cylinder renders "invisibility" to the transfer of angular momentum.

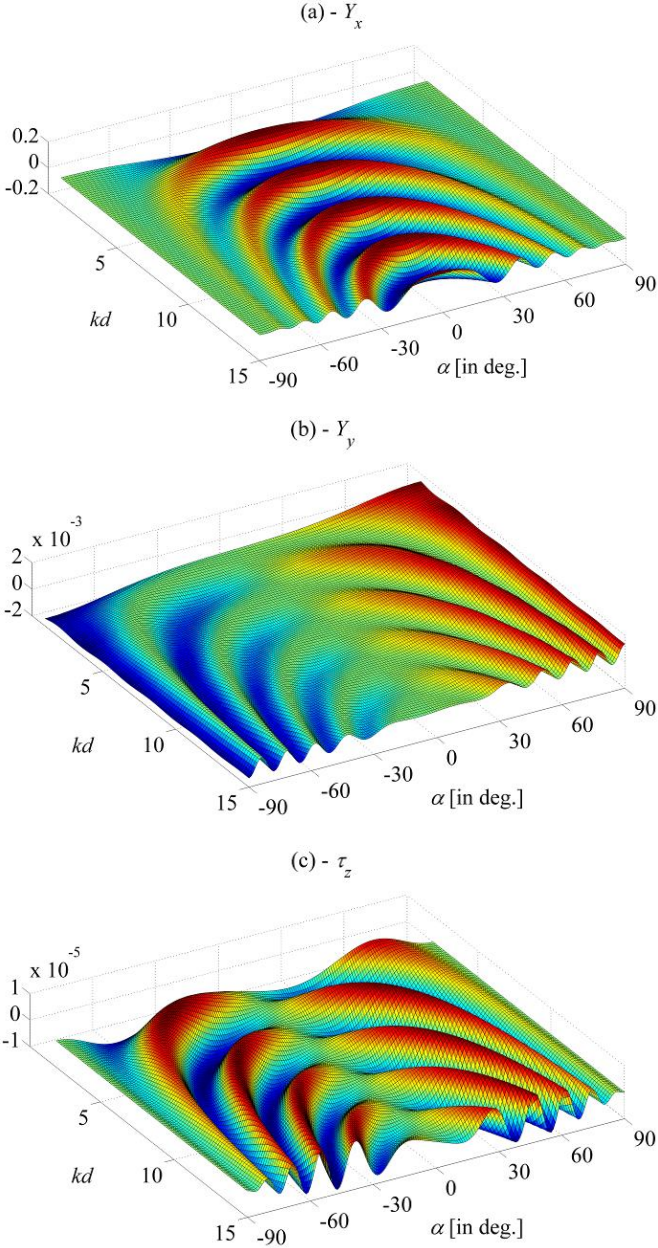

FIG. 3. Panels (a) and (b) show the plots for the longitudinal and transversal force functions, while panel (c) corresponds to the plot for the axial torque function, respectively, versus $kd$ and $\alpha$ for a small (subwavelength) viscous circular cylinder located near a boundary at $ka = 0.1$.

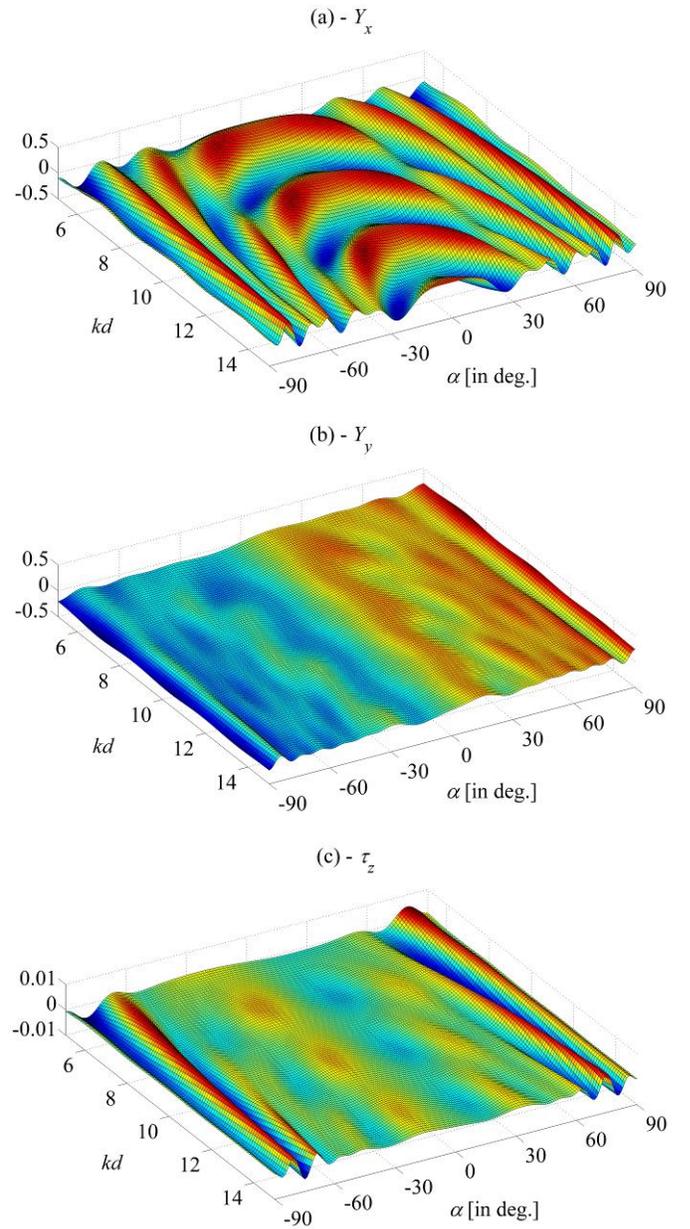

FIG. 4. The same as in Fig. 3 but $ka = 5$.

Increasing $ka$ (= 5) is considered now, where the related plots for the longitudinal and transversal components of the force functions are shown in panels (a) and (b) of Fig. 4, respectively. As $ka$ increased, $Y_x$ is about 2.5 times larger than its counterpart for the Rayleigh cylinder shown in panel (a) of reverses its spinning around its center of mass depending on the values of $kd$ and $\alpha$, as $ka$ increases. Moreover, $\tau_z(-\alpha) = -\tau_z(\alpha)$, where $\tau_z = 0$ at $\alpha = 0°$ as required by symmetry.

Fig. 3. The plot of $Y_x$ shows that both positive and negative values can be predicted, which anticipates the generation of repulsive and attractive longitudinal forces, contingent upon the values of $kd$ and $\alpha$, as $ka$ increases. The symmetry with respect to the axis $\alpha = 0°$ is also preserved such that $Y_x(-\alpha) = Y_x(\alpha)$. The plot of $Y_y$ displayed in panel (b), is also about 250 times larger than the plot shown in panel (b) of Fig. 3 for the Rayleigh viscous fluid cylinder. Similarly to panel (b) of Fig. 3, $Y_y(-\alpha) = -Y_y(\alpha)$, where $Y_y = 0$ at $\alpha = 0°$ as required by symmetry. Panel (c) shows the plot for the axial radiation torque function, which is also about 1000 times larger than its counterpart, as shown in panel (c) of Fig. 3. The oscillations of the axial radiation torque function are less emphasized, nonetheless, positive and negative values of $\tau_z$ are predicted numerically, which suggests that the viscous fluid cylinder

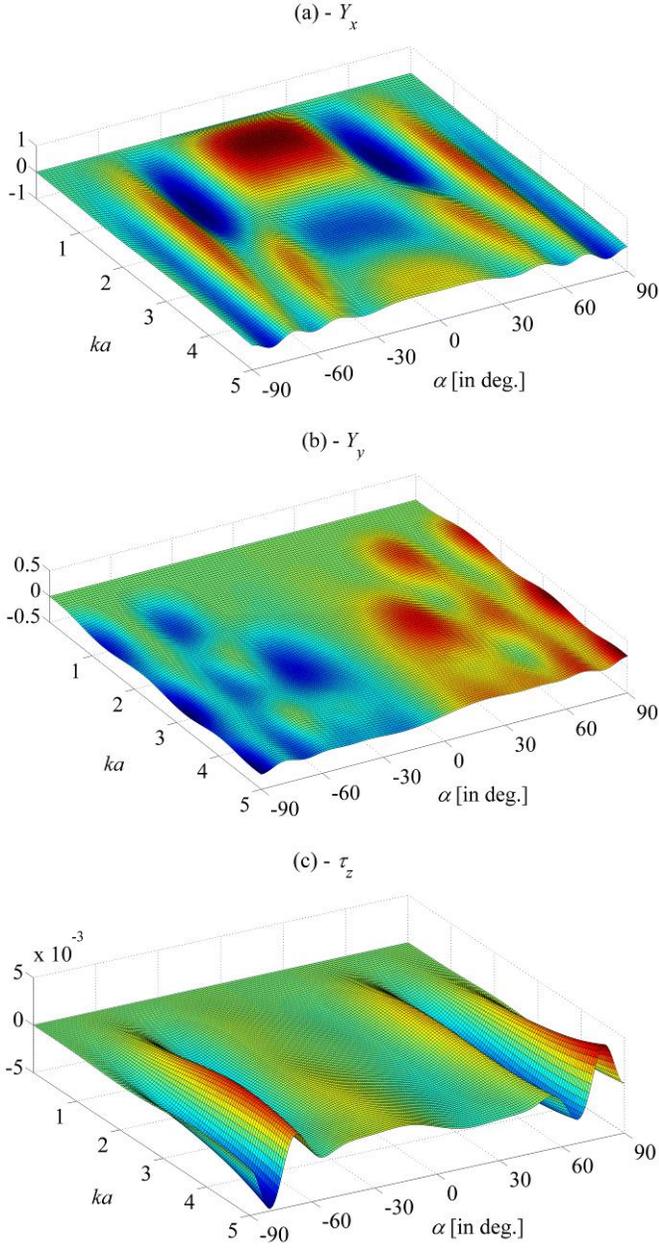

FIG. 5. Panels (a) and (b) exhibit the plots for the longitudinal and transversal force functions, while panel (c) shows to the plot for the axial torque function, respectively, versus $ka$ and $\alpha$ for a viscous circular cylinder located near a boundary at $kd = 5.5$.

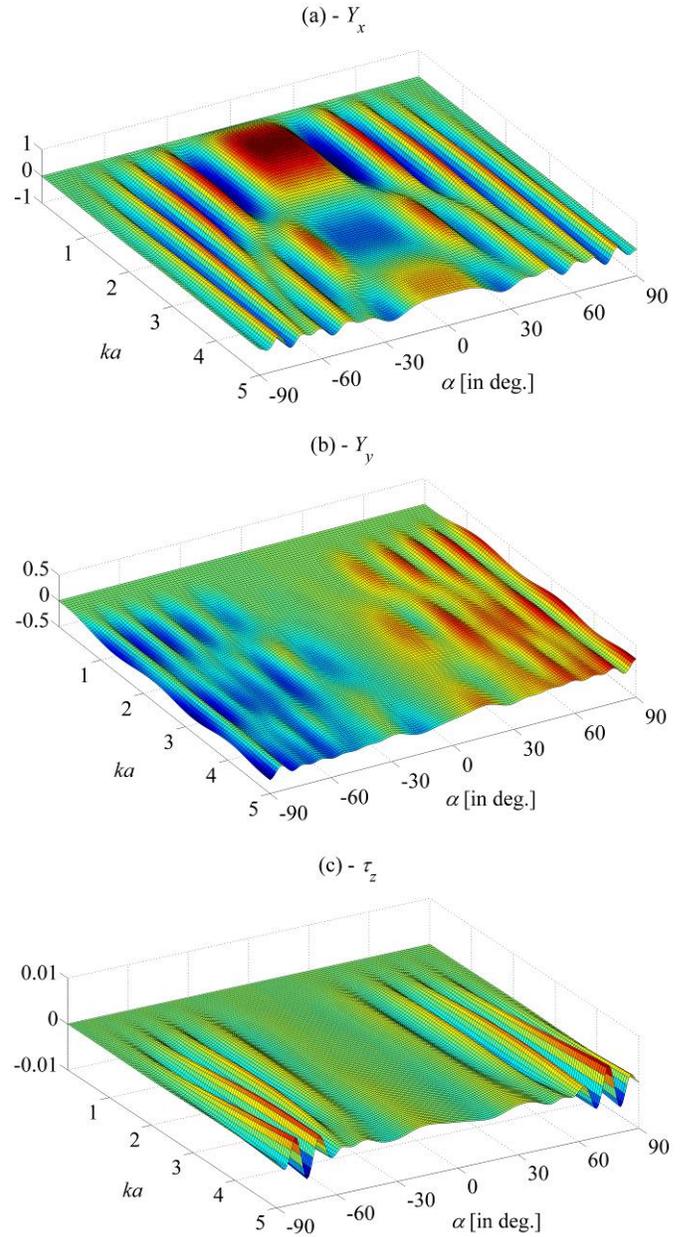

FIG. 6. The same as in Fig. 5 but $kd = 15$.

Another example of interest is to vary the dimensionless size parameter and angle of incidence, respectively, in the chosen ranges $0 < ka \leq 5$ and $-90° \leq \alpha \leq 90°$, for fixed values of the non-dimensional particle-boundary distance $kd$. Panels (a)-(c) of Fig. 5 exhibit the results for the longitudinal and transversal force functions, in addition to the axial torque function, for a lossy fluid cylinder located at $kd = 5.5$. In all the plots, the amplitudes of the force and torque functions increase as $ka$ increases, which is expected since at higher frequencies, the scattering and absorption increase, which both contribute to the force [23,43] and torque [14]. Panels (a)-(c) also show that the symmetry of the function $Y_x(-\alpha) = Y_x(\alpha)$ as well as the asymmetry of the functions $Y_y(-\alpha) = -Y_y(\alpha)$ and $\tau_z(-\alpha) = -\tau_z(\alpha)$ are preserved.

As $kd$ increases (= 15), the computed force and torque functions shown in panels (a)-(c) of Fig. 6 display noticeable undulations at a large but fixed $ka$ while $\alpha$ varies. At a larger particle-boundary dimensionless distance (i.e., $kd = 15$) the established (quasi)standing wave field generated between the object and the flat wall at a fixed $ka$ value possesses more nodes and antinodes, which affects the amplitudes of the radiation force and torque functions.

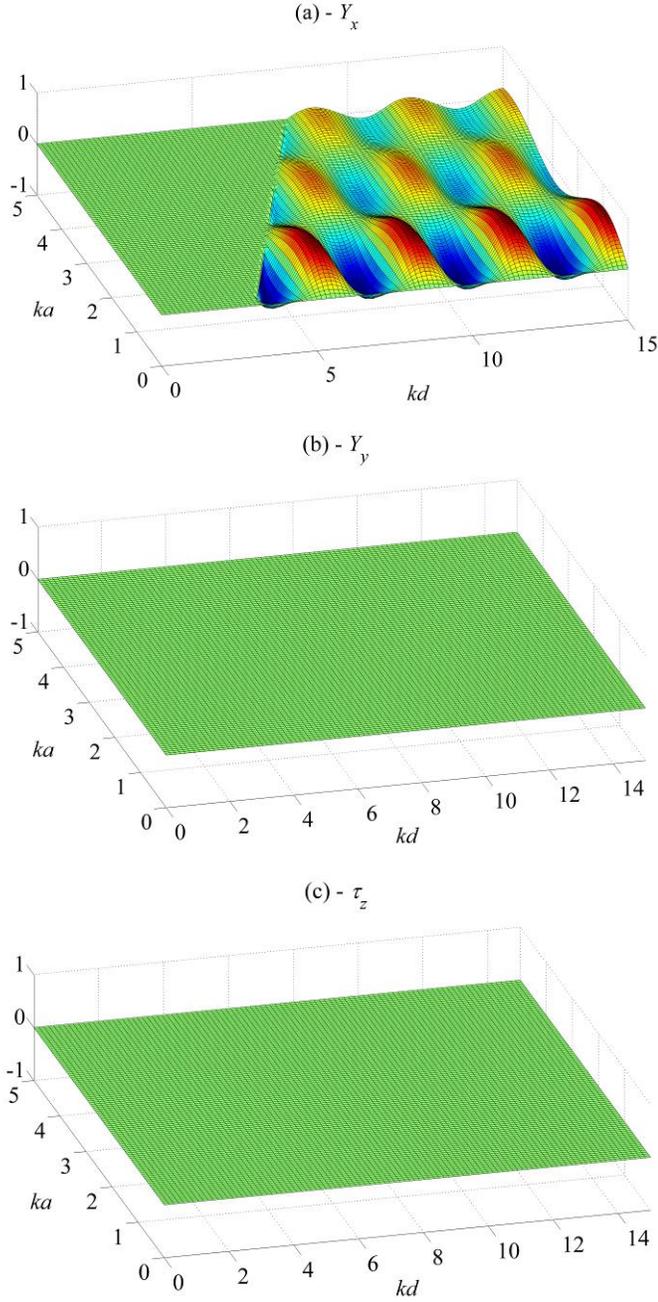

FIG. 7. Panels (a) and (b) show the results for the longitudinal and transversal force functions, while panel (c) corresponds to the plot for the axial torque function, respectively, versus $ka$ and $kd$ for a viscous circular cylinder located near a boundary at $\alpha = 0°$.

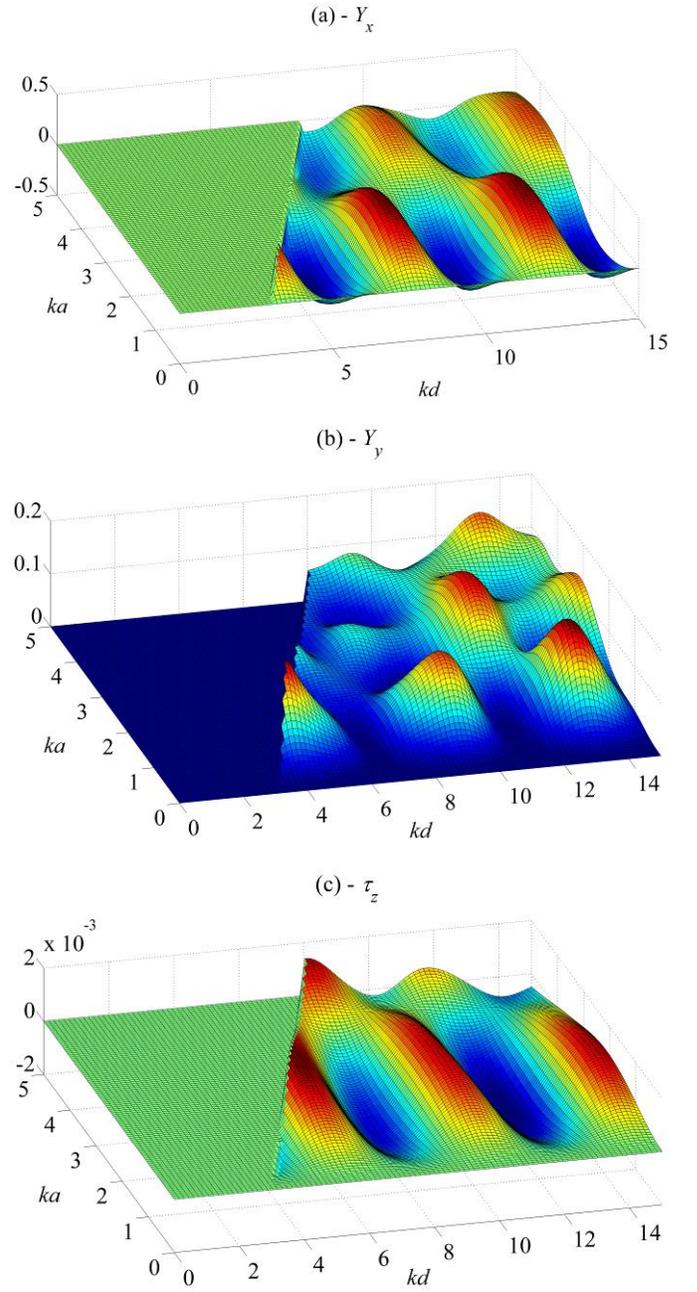

FIG. 8. The same as in Fig. 7 but $\alpha = 45°$.

The effect of varying the size parameter and the dimensionless distance, respectively, in the chosen ranges $0 < ka \leq 5$ and $ka < kd \leq 15$ is now considered for fixed values of the angle of incidence $\alpha$. Practically, for any physical situation, the constraint $kd > ka$ should be always satisfied, indicating that the distance particle-wall should be larger than the radius of the cylinder. After imposing this condition, the radiation force and torque functions are computed for $\alpha = 0°$, and the corresponding results are displayed, respectively, in Fig. 7. At this angle, only the longitudinal radiation force function is nonzero, as $Y_y$ and $\tau_z$ vanish as required by symmetry. Panel (a) of Fig. 7 shows that $Y_x$ oscillates between positive and negative values depending on the choice of $kd$ and $ka$, suggesting the possibility of pushing and pulling the cylindrical particle away or toward the boundary.

Increasing the angle of incidence to $\alpha = 45°$ is also considered, and the related computations for the radiation force and torque functions are displayed, respectively, in panels (a)-(c) of Fig. 8. As $\alpha$ deviates from zero, $Y_x$ displays smaller amplitudes compared to those shown in panel (a) of Fig. 7, while $Y_y$ is nonzero now. Panel (c) shows also that $\tau_z \neq 0$, and varies between + and − values as $kd$ and $ka$ vary. This also suggests that the viscous fluid cylinder can spin around its center of mass (counter-clockwise or clockwise) contingent upon the sign of $\tau_z$. At the crossing point where $\tau_z$ vanishes, the particle becomes unresponsive to the transferal of the angular momentum carried by the incident waves, yielding "rotational invisibility". Notice that the angular velocity of the cylinder can be determined (following the steps described in Section 4 of [14]) using the rotational friction coefficient [49] for a permeable cylinder.

## 4. Conclusion

This work presented a comprehensive analysis for the radiation force and radiation torque on a particle of arbitrary geometrical cross-section located nearby a boundary in 2D. Closed-form partial-wave expansions for the longitudinal and transversal radiation force functions, in addition to the axial radiation torque function (which remain suitable for any particle of arbitrary geometry in 2D) are derived, stemming from the integration of the radiation stress tensor and its moment. The analysis is also substantiated with computations for a circular viscous liquid cylinder immersed in a non-viscous fluid, and subjected to incident plane travelling waves with arbitrary incidence. The coupling effects of the scattered waves from the particle and its image, and those reflected by the flat rigid boundary are taken into account while deriving the mathematical expansions for the force and torque components. The multipole modal expansion method in cylindrical coordinates, the method of images, in addition to the translational addition theorem of cylindrical wave functions are used. Systematic examples are considered where a series of undulatory/wiggly behaviors are manifested in the plots of the radiation force and torque functions. Depending on the particle-boundary dimensionless distance $kd$, the incidence angle $\alpha$ as well as the dimensionless size parameter $ka$, situations have been predicted where the particle yields total force and torque neutrality depending on the choice of the parameters $\alpha$, $ka$ and $kd$. The present formalism finds useful applications in bio-acoustofluidics, particle manipulation, rotation and dynamics for a particle near a chamber wall. Other applications dealing with the interaction of acoustical waves on objects in underwater/ocean acoustics, sonochemistry, reconfigurable metafluids and liquid crystals may benefit from the results of the present investigation. The expressions for the force and torque components can serve to predict similar behaviors for a 3D particle of arbitrary shape (or spherical form) nearby a boundary, and this analysis should assist in developing appropriate analytical modeling using the multipole expansion method, the method of images and the addition theorem in spherical coordinates. Notice that recent advances [50-56] have considered the cases of non-spherical objects (i.e., spheroids) in Bessel beams, and the scope of such analyses can be extended to consider the interaction of acoustical waves with non-spherical particles nearby a boundary.